\documentclass[aps,prc,twocolumn,showpacs,superscriptaddress,nofootinbib,nolongbibliography,10pt]{revtex4-2}  
\usepackage{changes} 
\usepackage[para,online,flushleft]{threeparttable}
\usepackage{graphicx}  
\usepackage{dcolumn}   
\usepackage{bm}        
\usepackage{amssymb}   
\usepackage{amssymb,amsmath,amsfonts} 
\usepackage[breaklinks=true,debug=true]{hyperref}
\usepackage{braket}    
\usepackage{multirow}
\usepackage{adjustbox}
\usepackage{gnuplottex}
\usepackage[T1]{fontenc}
\usepackage[utf8]{inputenc}
\usepackage{color}
\usepackage{dsfont}
\usepackage{tikz}
\usepackage{pgfplots}
\usepgfplotslibrary{fillbetween}
\usetikzlibrary{spy,shadows}
\usetikzlibrary{backgrounds}
\usetikzlibrary {arrows.meta}
\usetikzlibrary{calc,patterns,angles,quotes}

\usepackage{empheq}

\usepgfplotslibrary{groupplots}
\pgfplotsset{compat=newest}
\usepgfplotslibrary{units}

\newcolumntype{M}[1]{>{\centering\arraybackslash}m{#1}}

\begin{document}


\title{Refining the isovector component of the Woods-Saxon potential}

\author{L.~Xayavong} 
\email{xayavong.latsamy@yonsei.ac.kr}
\affiliation{Department of Physics, Yonsei University, Seoul 03722, South Korea}

\author{Y.~Lim}
\email{ylim@yonsei.ac.kr}
\affiliation{Department of Physics, Yonsei University, Seoul 03722, South Korea}

\author{N.~A.~Smirnova} 
\email{smirnova@lp2ib.in2p3.fr}
\affiliation{LP2IB (CNRS/IN2P3-Universit\'e de Bordeaux), 33170 Gradignan cedex, France}

\author{G.~Nam}
\email{namgh@yonsei.ac.kr}
\affiliation{Department of Physics, Yonsei University, Seoul 03722, South Korea}

\vskip 0.25cm  
\date{\today}

\begin{abstract} 

We investigate the isovector component in the phenomenological mean field model of nuclei. 
Lane's isospin dependence, initially proposed for the nuclear optical potential, is reexamined within the context of bound states using the Woods-Saxon potential. 
We demonstrate that the original parametrization can be reexpressed in terms of parameters associated with the compound nucleus, enhancing its suitability for bound states. 
Comparisons with the conventional symmetry term are performed to assess how well each approach fits experimental data on single-particle/hole energies and reproduces charge-radius systematics. Our results indicate that Lane's formula provides better accuracy compared with the traditional approach to the nuclear potential.
Additionally, we find that the isovector component of the nuclear potential favors a surface-peaked form factor, especially one described by the first derivative of the Fermi like function  divided by the radial coordinate. This consideration is crucial for open-shell nuclei where 
Woods-Saxon eigenfunctions serve as a realistic basis for other many-body methods. 
Our findings also enable discrimination among various shell-model calculations of the isospin-symmetry breaking correction to superallowed $0^+\rightarrow0^+$ nuclear $\beta$ decays [I.~S. Towner and J.~C. Hardy, Phys. Rev. C {\bf 77}, 025501 (2008)]. This disparity currently constitutes the main source of theoretical uncertainty in subsequent tests of the standard model. 

\end{abstract}

\maketitle

\section{Introduction}\label{sec1}

The phenomenological Woods-Saxon (WS) potential represents a suitable choice to generate realistic radial wave functions. Besides the applicability to non correlated systems, these eigenfunctions also serve as an efficient basis for modern nuclear many-body methods, particularly in limited configuration spaces. Various WS parameter sets are available, including those discussed in Refs.\,\cite{SWV,WS1,WS2,WS3}. These parameter sets, however, were typically obtained via global fittings to specific nuclear properties and are generally appropriate only within the regions of the nuclear chart where they were fitted. Their accuracy tends to diminish as one moves further from the valley of stability. Therefore, optimizing WS parameters is an essential process toward the future large-scale nuclear structure calculations. 

Among various applications of a WS potential, those related to the weak-interaction studies on nuclei require particular precision and control of the potential parameters. The most well known and important case of such calculations is related to the theoretical analysis of the nuclear Fermi $\beta $ decay.
Indeed, superallowed $0^+\rightarrow 0^+$ nuclear $\beta$ decay of isotriplets ($T=1$) offers an excellent tool for testing fundamental symmetries and the low-energy structure of the electroweak interactions underlying the standard model\,\cite{HaTo2020,RevModPhys.78.991,GONZALEZALONSO2019165}. In particular, it enables an experimental extraction of the top-left element, $V_{ud}$, of the Cabibbo–Kobayashi–Maskawa (CKM) quark-mixing matrix\,\cite{Cabibbo1963,KobayashiMaskawa1973}. The master formula for this semileptonic process is written as
\begin{equation}\label{master}
    Ft=ft(1+\delta_R')(1-\delta_C+\delta_{NS}) = \frac{K}{2G_F^2V_{ud}^2(1+\Delta_R^V)}, 
\end{equation}
where $K$ is a combination of fundamental constants\,\cite{OrBr1985} and $G_F$ is the Fermi coupling constant\,\cite{PhysRevLett.106.041803,10.21468/SciPostPhysProc.5.016}. The experimental inputs in Eq.\,\eqref{master} comprise the statistical rate function $(f)$\,\cite{PhysRevC.91.015501,PhysRevC.109.045501} and partial half-life $(t)$, collectively denoted as $ft$. The $ft$ values have been measured with sub-percent precision for 15 cases, spanning the mass range from $^{10}$C to $^{74}$Rb\,\cite{HaTo2020}. The radiative correction is divided into three terms: $\Delta_R^V$ represents the nucleus-independent component, $\delta_R'$ depends only on the atomic number and the decay $Q$-value, and $\delta_{NS}$ represents the nuclear structure-dependent component. The complete detailed formalism and recent improvements for the radiative correction terms can be found in Refs.\,\cite{PhysRevC.107.035503,PhysRevD.100.013001,TOWNER1992478,PhysRevD.108.053003,particles4040034,PhysRevLett.121.241804}. Another nuclear structure input that is the central focus of our current research is $\delta_C$, the isospin-symmetry breaking correction. 

The conserved vector current (CVC) hypothesis implies that the vector coupling constant, $G_V$, which is related to $V_{ud}$ by $G_V^2=G_F^2V_{ud}^2$, should not be renormalized in the nuclear medium. 
Therefore, the validity of CVC can be tested through the constancy of the corrected $Ft$ across various nuclei, as indicated by formula\,\eqref{master}.

Furthermore, the fluctuation in $Ft$ as a function of the average inverse decay energy in light nuclei is useful for setting robust limits on the existence of induced scalar currents\,\cite{HaTo2020,RevModPhys.46.789}. Among the required theoretical inputs, $\delta_C$ generally exhibits strong variations when moving from one nucleus to another. This correction has been intensively studied using various theoretical approaches, ranging from simplified schematic models that incorporate Coulomb mixing with harmonic oscillator wave functions to first-principle methods such as the ab-initio no-core shell model in Refs.\,\cite{XaNa2018,XaNa2022,xayavong2022higherorder,Xa2017,Dam1969,TOWNER1977269,ToHa2008,PhysRevC.86.054316,Auerbach2009,Li2009,nocore,Lam2013,PhysRevC.92.055505,OrBr1985,OrBr1989x,OrBr1995,OrBr1989,MiSch2008,MiSch2009,Xthesis,AUERBACH2022122521,PhysRevLett.106.132502,Calik2013,PhysRevC.88.064318,particles4040038,GRINYER2010236,PhysRevC.109.044302}. However, among these approaches, only the phenomenological shell model yields reasonable agreement with CVC across a broad range of nuclei, as demonstrated in the comparative test in Ref.\,\cite{ToHa2010}. It has also been validated in certain cases with sensitive experimental tests, such as those conducted in Refs.\,\cite{PhysRevLett.112.102502,PhysRevLett.73.396,ToHa2008}. 

Since the latest survey by Hardy and Towner\,\cite{HaTo2020}, it has been suspected that the primary sources of errors in $V_{ud}^2$ originate from the experimental inputs and radiative correction terms.
When combined with data for $V_{ub}$\,\cite{PhysRevD.98.030001} and $V_{us}$\,\cite{Aoki2020}, the extracted $V_{ud}^2$ value shows a discrepancy with the unitarity condition of the CKM matrix by more than 2 standard deviations\,\cite{HaTo2020}.
However, before drawing any conclusions about the standard model and new physics, it is crucial to ensure that there are no undetected sources of errors in either the experimental or theoretical inputs listed above. Several ongoing experimental programs aimed at improving the precision of half-life and branching ratio measurements have been launched using
radioactive ion beam facilities worldwide\,\cite{acharya2023}.
These programs seek to enhance the precision for the 15 cases that have already reached a sub-percent precision level. In addition, the program includes other cases that are currently close to the required precision threshold but are not yet part of this ensemble. 

The WS potential plays an important role in the shell-model approach\,\cite{xayavong2022higherorder,XaNa2018,Xa2017,NaXa2018,ToHa2008,Lam2013}. Within this theoretical framework, $\delta_C$ can be decomposed at the leading order into two terms.
The first term, $\delta_{C1}$, accounts for the isospin mixing within the shell-model valence space induced by the isospin-nonconserving component of the effective Hamiltonian. Whereas the second term, $\delta_{C2}$, corrects for the radial mismatch between proton and neutron wave functions, effectively capturing the isospin mixing that extends beyond the model space. Generally, $\delta_{C2}$ is approximately one order of magnitude larger than $\delta_{C1}$, especially when the transition involves weakly bound nuclei\,\cite{xayavong2022higherorder,ToHa2008}. In this approach, it is crucial that the WS radial wave functions used for evaluating $\delta_{C2}$ accurately reproduce experimental data. These data typically include separation energies, which ensure accurate asymptotics, and charge radii, which provide a robust constraint on the potential geometry.
Currently, there is still some freedom in achieving this requirement. For instance, one can reproduce separation energies, assuming the validity of Koopman's theorem\,\cite{KOOPMANS1934104}, either by readjusting the overall depth of the volume term or by varying the strength of an additional surface-peaked term, as employed in Refs.\,\cite{ToHa2008,XaNa2018,xayavong2022higherorder}. 
As these refinement processes are separately applied for protons and neutrons, they strongly influence the nuclear isovector component and generally introduce an additional source of isospin-symmetry breaking into the WS potential. The procedure of readjusting the depth or the volume-term strength is essentially equivalent to adding an isovector term with the usual WS-potential form factor, 
\begin{equation}\label{f}
\displaystyle
f(r) = \frac{1}{1+\exp\left( \frac{r-R}{a_0}\right)}, 
\end{equation}
where $a_0$ is the surface diffuseness parameter and $R=r_0\times(A-1)^\frac{1}{3}$ with $r_0$ denoting the length parameter. Similarly, other process can be seen as adding an isovector term, within a surface-peaked form factor. 

The non-uniqueness of these refinement procedures constitutes a significant source of uncertainty for $\delta_{C2}$. More details on the uncertainty quantification for this correction can be found in Refs.\,\cite{ToHa2008,ToHa2002,XaNa2018}. 

The paper aims at achieving two primary goals. Firstly, we investigate the validity of Lane's formula\,\cite{LanePRL1962,Lane1962} for bound states of nucleons. To avoid ambiguity regarding the target nucleus, this formula will be re-expressed in terms of well-defined quantum numbers of the compound system. A comparison with the conventional asymmetry term\,\cite{XaNa2018,PhysRevC.108.064310} will also be performed. Secondly, we explore the spatial dependence of the nuclear isovector component. The different radial form factors mentioned in the previous paragraph will be investigated, with a particular focus on
the surface peaking observed in prior studies\,\cite{HUANG2023138293,DG,Ahmad_1976,AHMAD196991,PINKSTON1965641}. 
Comparative studies among these choices will be conducted based on the quality of parameter fittings, 
considering the single-particle and single-hole states in the vicinity of closed-shell nuclei, 
and the ability to reproduce the charge-radius systematics. 
The outcomes of these studies will have important potential implications for the shell-model calculations of $\delta_C$ and the subsequent tests of the standard model\,\cite{HaTo2020,ToHa2008}. 

\section{Nuclear isovector potential}\label{sec2} 

The WS potential is widely regarded as realistic for modeling effective nuclear interactions within the mean-field approach\,\cite{PhysRev.95.577,Hod1971,PhysRev.111.1147,PhysRev.109.429,Lane1962,SWV,PhysRevC.108.064310,XaNa2018,ToHa2008,WS1,WS2,WS3}. It is parametrized based on nuclear-force properties, incorporating Coulomb and spin-orbit terms to address the Coulomb repulsion among protons and ensure the correct sequences of traditional magic numbers, respectively. The Coulomb term has a well-known origin and is conventionally evaluated using the approximation of a uniformly charged sphere. More sophisticated approaches to this repulsive term are also available\,\cite{PhysRevC.108.064310} if greater precision is required. The spin-orbit term is evaluated using the Thomas precession\,\cite{Thomas} which implies a spatial dependence in the form of the first derivative of the volume-term form factor divided by the radial coordinate. Note that we assume the spherical symmetry throughout this paper. As an empirical refinement, the recent parametrizations\,\cite{ToHa2008,XaNa2018,PhysRevC.108.064310} tend to adopt a smaller length-parameter value for the spin-orbit term, as the two-body spin-orbit interaction exhibits a shorter range\,\cite{BohrMott}. 

The isovector term in the mean-field potential has multiple microscopic origins. It can be attributed to the charge-symmetry and charge-independent breaking components of nucleon-nucleon interactions\,\cite{DG,XaNa2022}. Additionally, Pauli's exclusion principle, which prohibits the occupation of a single-particle state by more than one identical nucleon, also causes an additional shift between proton and neutron levels. Conventionally, the nuclear isovector property of WS potential is described by the symmetry term\,\cite{WS1,WS2,WS3,XaNa2018,BohrMott}, often embedded within the volume (isoscalar) term as
\begin{equation}\label{sys}
    V = V_0\left( 1 \pm \kappa\frac{N-Z}{A} \right), 
\end{equation}
where the upper (lower) sign corresponds to proton (neutron). The parameters $V_0$ and $\kappa$ depend in general on the relative momentum between 
the incident nucleon and the target nucleus,
but do not depend on $Z$, $N$ and $A$. However, the momentum dependence can be neglected for low-energy bound states. As usual, $Z$, $N$, and $A$ represent the atomic, neutron, and mass number, respectively. The symmetry term\,\eqref{sys} is generally non-negligible, especially when the difference $(N-Z)$ becomes considerable. It destructively contributes to the total binding energy, as can be seen from the empirical mass formula\,\cite{MOLLER1988213}. This property is consistent with the experimental observation that the binding energy exhibits a maximum around the $N=Z$ nucleus\,\cite{AME2012,AME2020I,AME2020II} within a specific isobaric chain. 
Thus, in general a better understanding and modelization of the symmetry term may help to get more insight on the properties of the nuclear symmetry energy.

Despite its remarkable success, the parametrization in Eq.\,\eqref{sys} was considered to be a special case of a more fundamental expression that involves isospin dependence, specifically the one introduced by Lane\,\cite{LanePRL1962,Lane1962}. 
Lane's formula was originally proposed to represent the interaction of an unbound nucleon with a target nucleus. It is expressed as 
\begin{equation}\label{Lane}
    V = V_0\left( 1 - \frac{4\kappa}{A-1}\braket{\bm{t}\cdot\bm{T}_{A-1}} \right), 
\end{equation}
where $\bm{T}_{A-1}$ and $\bm{t}$ denote the isospin operators of the target nucleus and the incident nucleon, respectively. 

It should be noted that the factor $1/(A-1)$ is used in Eq.\,\eqref{Lane} instead of $1/A$ as in Eq.\,\eqref{sys}. Historically, this choice stems from the scenario involving unbound nucleons, where the potential is predominantly created by the target nucleus, without significant contribution from the incident nucleon. For consistency, we will substitute $1/(A-1)$ in Eq.\,\eqref{Lane} with $1/A$ for the remainder of this paper. This substitution would not pose any significant issue, as the introduction of these factors is not strictly based on a solid theoretical foundation. We have verified that the error caused by this difference is always smaller than fitting RMS errors even in light nuclei such as $^{12}$C. 

The isospin dependence\,\eqref{Lane} has several important implications for nuclear reaction studies\,\cite{Lane1962,LanePRL1962,Hod1971,ERRAMUSPE1967569,HODGSON1963352,PhysRevC.71.044601}. Lane\,\cite{Lane1962} has demonstrated in an elaborate manner that an averaged form of Eq.\,\eqref{Lane} corresponds to the conventional ansatz given in Eq.\,\eqref{sys}, which describes the asymmetry contribution in the target nucleus. This correspondence is also justified when the target nucleus's isospin aligns with the projection along the $z$ axis, which is generally the case for the ground state\footnote{The ground states with $T\ne |T_z|$ are observed only in odd-odd self-conjugated nuclei heavier than potassium. This phenomenon is known as \emph{isospin inversion}\,\cite{JANECKE200587,PhysRevC.66.024327}.}. That is, $\braket{\bm{t}\cdot\bm{T}_{A-1}}=t_zT_{z,A-1}=\mp\frac{1}{4}(N'-Z')$ where $N'$ and $Z'$ denote the neutron and proton numbers of the target nucleus, respectively. Note that the $-(+)$ sign is for the operation to neutrons\,(protons). 
However, it is remarkable that while Eq.\,\eqref{Lane} appears to be well-founded, its applicability for bound nucleons is ambiguous due to strong coupling in the potential, which prevents the target nucleus from being regarded as an isolated system. Additionally, both Eqs.\,\eqref{sys} and \eqref{Lane} assume an identical form factor for their isoscalar and isovector parts, which could be oversimplified, especially when applying to open-shell nuclei. More fundamental expressions will be explored in Sec.\,\ref{sec5}. There is also evidence of an isovector component in the spin-orbit term\,\cite{XU2013247,PhysRevC.92.024306}; this effect, however, is expected to be 
smaller, particularly 
for nuclei around the $N=Z$ line, and will not be considered in this study. 

\section{Derivation of Lane's formula for bound states}\label{sec3} 

In principle, the potential describing the interaction between two particles can include all symmetry-preserving terms, including those involving the relative position and momentum operators, as well as the individual spins and isospins. Lane's formula\,\eqref{Lane} also preserves the symmetries of a system comprising an incident nucleon and a target nucleus. Therefore, it should be considered fundamental, at least in the context of this two-body approach. To derive the corresponding formula suitable for bound states, we decompose the total isospin operator of the compound system into 
\begin{equation}\label{e1}
\bm{T}=\bm{T}_{A-1} + \bm{t}, 
\end{equation}
with $\bm{T}_{A-1}$ representing the isospin operator of the virtual $(A-1)$-nucleon system, which is now not definite due to strong coupling with the incident nucleon to form bound state. We employ the isospin convention of $t=\frac{1}{2}$ for neutrons and $-\frac{1}{2}$ for protons. 

\begin{center}
\begin{figure}[ht!]
\begin{tikzpicture}
  \begin{axis}[ticks=none, ymin=-20, ymax=20, xmax=5, xmin=-1.5, xticklabel=\empty, yticklabel=\empty, axis lines* = middle, 
  axis line style={line width=1pt}, 
   xlabel=$xy$, ylabel=$z$, label style={at={(ticklabel cs:1.1)}}, ]
   \draw[-{Stealth}, very thick, red] (axis cs:0,0) -- node[above]{$\bm{T}$} (axis cs:4,15);
   \coordinate (o) at (0,0);
   \coordinate (z1) at (0.28,1);
   \coordinate (y) at (0,1);
   \draw pic[draw, "$\theta$", -{Stealth}, angle eccentricity=1.2, angle radius=1cm] {angle=z1--o--y};
   \draw[-{Stealth}, very thick, blue] (axis cs:0,0) -- node[above,sloped]{Neutron} (axis cs:0,10);
   \draw[{Stealth}-{Stealth}, thick] (axis cs:-0.7,0) -- node[left]{$T_z$} (axis cs:-0.7,15);
   \draw[thick, dashed] (axis cs:0,15) -- (axis cs:4,15);
   \draw[thick, dashed] (axis cs:4,0) -- (axis cs:4,15); 
   \draw[-{Stealth}, very thick, red] (axis cs:0,0) -- node[above]{$\bm{T}$} (axis cs:4,-15);
   \coordinate (o1) at (0,0);
   \coordinate (z2) at (0.28,-1);
   \coordinate (y1) at (0,-1);
   \draw pic[draw, "$\theta$", {Stealth}-, angle eccentricity=1.2, angle radius=1cm] {angle=y1--o1--z2};
   \draw[-{Stealth}, very thick, blue] (axis cs:0,0) -- node[above,sloped,rotate=180]{Proton} (axis cs:0,-10);
   \draw[{Stealth}-{Stealth}, thick] (axis cs:-0.7,0) -- node[left]{$T_z$} (axis cs:-0.7,-15);
   \draw[thick, dashed] (axis cs:0,-15) -- (axis cs:4,-15);
   \draw[thick, dashed] (axis cs:4,0) -- (axis cs:4,-15); 
   \node[] at (1.8,18) {Neutron-rich nuclei} ;
   \node[] at (1.8,-18) {Proton-rich nuclei} ;
  \end{axis}
\end{tikzpicture}
\caption{\label{figxx}(Color online). Illustration of the isospin alignments relevant for the evaluation of $\braket{\bm{t}\cdot\bm{T}}$. 
} 
\end{figure}
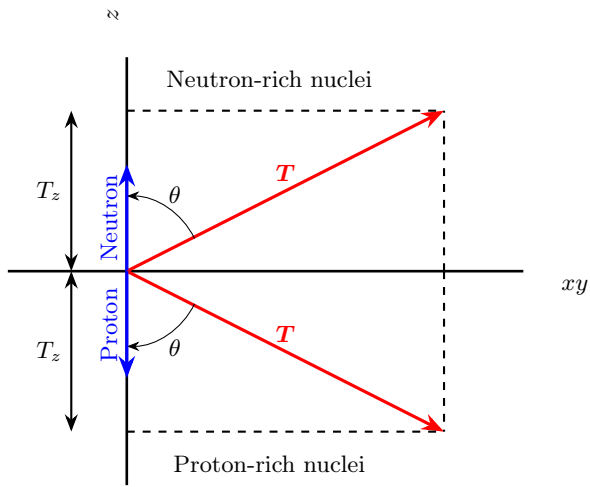
\end{center}

Multiplying both sides of Eq.\,\eqref{e1} by the operator $\bm{t}$, we arrive at 
\begin{equation}\label{e2}
    \braket{\bm{t}\cdot\bm{T}_{A-1}} = \braket{\bm{t}\cdot\bm{T}} - \braket{\bm{t}^2}, 
\end{equation}
where $\braket{\bm{t}^2}=t(t+1)=\frac{3}{4}$. The first term on the right-hand side of Eq.\,\eqref{e2} can be expressed as $\braket{\bm{t}\cdot\bm{T}}=t T\cos(\theta)$, with $\theta$ representing the relative angle. Since $\bm{t}$ can only be aligned (neutrons) or anti-aligned (protons) with the $z$ axis, the projection of $\bm{T}$ onto $\bm{t}$ is simply proportional to its projection onto the $z$ axis (see Fig.\,\ref{figxx}). Consequently, the magnitude of this term does not depend on the absolute $T$ value but is solely determined by its $z$-axis component $T_z$, specifically $\braket{\bm{t}\cdot\bm{T}}=t_zT_z=\mp\frac{1}{4}(N-Z)$. The sign of $\braket{\bm{t}\cdot\bm{T}}$ depends on whether $\bm{t}$ is aligned or anti-aligned, and whether the compound nucleus is neutron-rich or proton-rich. For better clarity, all relevant isospin configurations are shown in Fig.\,\ref{figxx}. From these results, Eq.\,\eqref{e2} is rewritten as  
\begin{equation}\label{e3}
    \braket{\bm{t}\cdot\bm{T}_{A-1}} = \mp \frac{(N-Z)}{4} - \frac{3}{4}. 
\end{equation} 

\begin{figure*}[ht!]
\begin{center}
\includegraphics[width=\textwidth]{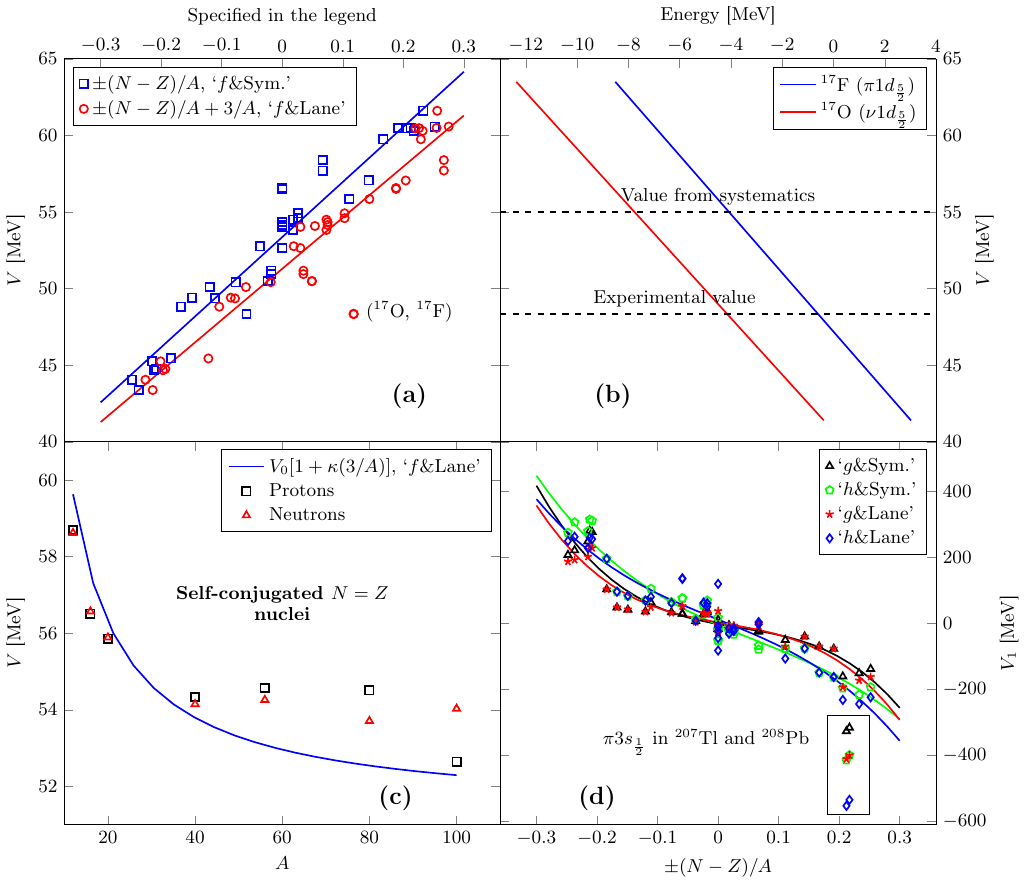}
    \caption{\label{fig1}(Color online) Panel (a) displays the $V$ values, 
    plotted against $\pm(N-Z)/A$ and $\pm(N-Z)/A+3/A$. 
    These correspond to the `$f$\&Sym.' and `$f$\&Lane' models in Table\,\ref{tab3}, respectively. 
    The outlayers, corresponding to $^{17}$O ($\nu1d_\frac{5}{2}$) and $^{17}$F ($\pi1d_\frac{5}{2}$), are excluded from the linear regressions. 
    Panel (b) shows the correlation between $1d_\frac{5}{2}$-state energies and $V$ for the mirror pair $^{17}$O/$^{17}$F. 
    The experimentally deduced and desired values of $V$ for these nuclei are indicated with horizontal dashed lines. 
    Panel (c) displays the dependence of $V$ on $A$ in self-conjugated nuclei. The solid curves represent extrapolated values.
    Note that the $V$ values shown in Panels (a), (b), and (c) are evaluated using the $f(r)$ form factor for both isoscalar and isovector parts. 
    Panel (d) illustrates the $V_1$ values, evaluated with the different models described in Table\,\ref{tab3}, as a function of $\pm(N-Z)/A$. The large scattered points in this plot, corresponding to the $\pi3s_\frac{1}{2}$ states in $^{207}$Tl and $^{208}$Pb, are also excluded from the regression analysis. }
\end{center}
\end{figure*}

Substituting Eq.\,\eqref{e3} into Eq.\,\eqref{Lane}, the original Lane's formula can be re-expressed in terms of the compound nucleus parameters. The final result is as follows
\begin{equation}\label{Lane1}
    V = V_0\left[ 1 \pm \kappa\frac{(N-Z)}{A} + \kappa\frac{3}{A} \right]. 
\end{equation} 
Note again that, for consistency, we use $A$ in the denominators instead of $A-1$ as in the original formulation. 

It is interesting to emphasize that the derived formula\,\eqref{Lane1} includes an additional term (the third term on the right-hand side) that makes it more attractive than Eq.\,\eqref{sys} when considering identical values for $V_0$ and $\kappa$. This indicates that the averaged version of the formula\,\eqref{Lane1} found in Ref.\,\cite{Lane1962} is not fully valid for bound states. This additional term is expected to have a strong impact in light nuclei as it is inversely proportional to the mass number, especially in $N\approx Z$ nuclei where the symmetry counterpart (the second term in Eq.\,\eqref{Lane1}) becomes negligible. 
We notice that a similar derivation has been conducted in Ref.\,\cite{SWV}. However, the resulting expressions presented therein appear to contain errors, except for the $N=Z$ case. 

Isospin dependence is not explicitly presented in the self-consistent Hartree-Fock (HF) mean field when using the effective zero-range Skyrme interaction. 
However, it was shown that the symmetry-like structure\,\eqref{sys} can be approximately recovered if the factor $(N-Z)/A$ is replaced with the respective matter densities, such as $[\rho_n(r)-\rho_p(r)]/[\rho_n(r)+\rho_p(r)]$\,\cite{DG}. Conceptually, this discovery establishes an important bridge between the phenomenological WS model and self-consistent HF mean field. However, it is likely that many aspects of the derivation in Ref.\,\cite{DG} were oversimplified. 
\begin{figure}[ht!]
\begin{center}
\begin{tikzpicture}
\begin{groupplot}[width=\linewidth, height=0.44\textwidth, group style={group size=1 by 1,xlabels at=edge bottom},
ylabel={Form factors}, xlabel={$r$~[fm]}, legend style={/tikz/every even column/.append style={column sep=0.4cm}, at={(0.98,0.98)}, anchor=north east,legend columns=1}, legend cell align={left}, grid=both ] 
    \nextgroupplot[xmin=0, xmax=6, ymin=-0.1, ymax=1]
    \pgfmathsetmacro{\R}{3} 
    \pgfmathsetmacro{\a}{0.662} 
    \addplot[blue, thick, domain=0.1:7, samples=20] { 1/(1+exp((x-\R)/\a)) } ; 
    \addlegendentry{$f(r)$}
    \addplot[black, thick, domain=0.1:7, samples=40] { \a^2*( exp((x-\R)/\a)/(1+exp((x-\R)/\a)^2 ) )^2 } ; 
    \addlegendentry{$h(r)$} 
    \addplot[red, thick, domain=0.1:7, samples=50] { 2*(1/(\a*x))*exp((x-\R)/\a)/(1+exp((x-\R)/\a) )^2 } ; 
    \addlegendentry{$g(r)$} 
    \draw[thick, dashed] (3,-0.2)-- node[below,sloped,pos=0.68] {$R=3$~fm} (3,1.2) ;
\end{groupplot}
\end{tikzpicture}
\caption{\label{fig0}(Color online). Comparison of the form factors employed in this study. The vertical line indicates the radius parameter, $R=3$\,fm. The surface diffuseness parameter, $a_0$, is set to 0.662\,fm\,\cite{XaNa2018}.}
\end{center} 
\end{figure}
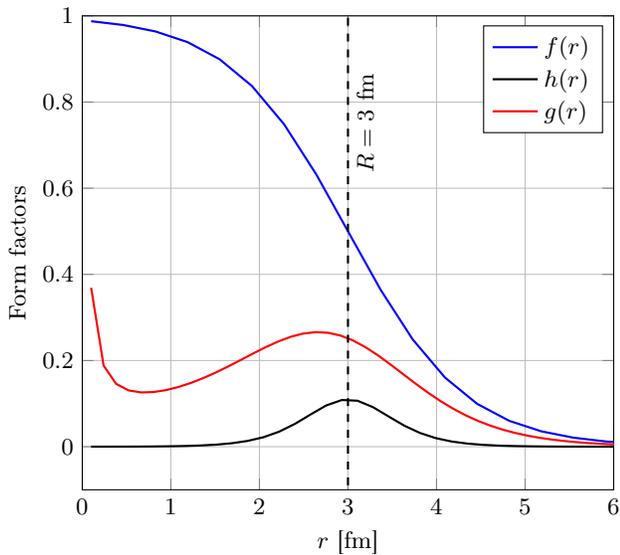

\section{Comparative study between Lane's formula and the symmetry term}\label{sec4} 

The difference between Eqs.\,\eqref{sys} and \eqref{Lane1} prompts us to conduct a comparative study 
of their accuracy and draw a discrimination if necessary. 
Most seminal works contributing to the development of the phenomenological nuclear mean-field theory were conducted within the optical model framework\,\cite{Hod1971,HODGSON1963352,Lane1962,LanePRL1962,VARNER199157,LIPPERHEIDE196697}. Applications to nuclear structure are much less intensive, as the majority of nuclei are significantly influenced by correlations beyond the mean-field approach. Despite this limitation, our present study focuses exclusively on bound states, in particular the one-particle and one-hole configurations in the vicinity of the doubly magic cores\,\cite{SWV}. Additionally, we include the single-particle states observed in closed-(sub)shell nuclei\,\cite{Vautherin,Bespalova2007}, whose energies are generally obtained as weighted averages to account for the fragmentation of spectroscopic strengths\footnote{We acknowledge that the energies obtained in this manner only represent the centroid. However, they should be appropriate for fitting purposes, where the outcome is sensitive to global trends rather than local fluctuations. }.

We vary $V$ for each selected state to match the WS eigenvalues with the experimental data. 
The other components of WS potential, including the Coulomb and spin-orbit terms, are kept fixed at their standard values\,\cite{XaNa2018}. 
As a center-of-mass correction, we replace the nucleon mass in the kinetic term of the radial Schrödinger's equation with a reduced mass, as described in Ref.\,\cite{SWV,XaNa2018}. This approach is asymptotically correct as the virtual $(A-1)$-nucleon system becomes isolated and resembles a point-like (structureless) particle at large separation. It is also assumed that the mass difference between the proton and neutron is negligible. 
The resulting $V$ values for the individual states, to be used to test Eqs.\,\eqref{sys} and \eqref{Lane1}, are listed in Table\,\ref{tab1}. 

To facilitate our references, we use unique labels to represent specific models for the isoscalar and isovector components of the WS potential. For instance, we use the labels `$f$\&Sym.' and `$f$\&Lane' for the models in Eqs.\,\eqref{sys} and \eqref{Lane1}, respectively. More complete details, including the other models considered in the subsequent section, are given in Table\,\ref{tab3}. 

It is interesting to note that both `$f$\&Sym.' and `$f$\&Lane' are, by construction, invariant under the exchange of $Z$ and $N$ (charge symmetry). This implies that the $V$ values obtained for a pair of mirror nuclei should be identical. The nuclear charge-symmetry breaking component, which is not explicitly accounted for in WS potential, is generally much weaker\,\cite{DG} and can typically be absorbed into the Coulomb term. Our numerical results provided in Table\,\ref{tab1} are very well consistent with this property. 


Panel (a) of Fig.\,\ref{fig1} displays the $V$ values as a function of $\pm(N-Z)/A$ and $\pm(N-Z)/A+3/A$, which are the independent variables for the `$f$\&Sym.' and `$f$\&Lane' models, respectively. In the plot against $\pm(N-Z)/A+3/A$, we observe two distinct points that deviate from the global trend by more than 5\,MeV. These points correspond to $^{17}$O ($\nu1d_\frac{5}{2}$) and $^{17}$F ($\pi1d_\frac{5}{2}$). In contrast, the plot of the same data against $\pm(N-Z)/A$ does not exhibit any unusual behavior. At first glance, this anomaly seems to be attributed to the additional term in `$f$\&Lane', specifically the one proportional to $3/A$, which is not present in `$f$\&Sym.'. However, if this term were solely responsible, similar effects would also be observed in other light nuclei.
\begin{figure*}[ht!]
  \begin{center}
    \includegraphics[]{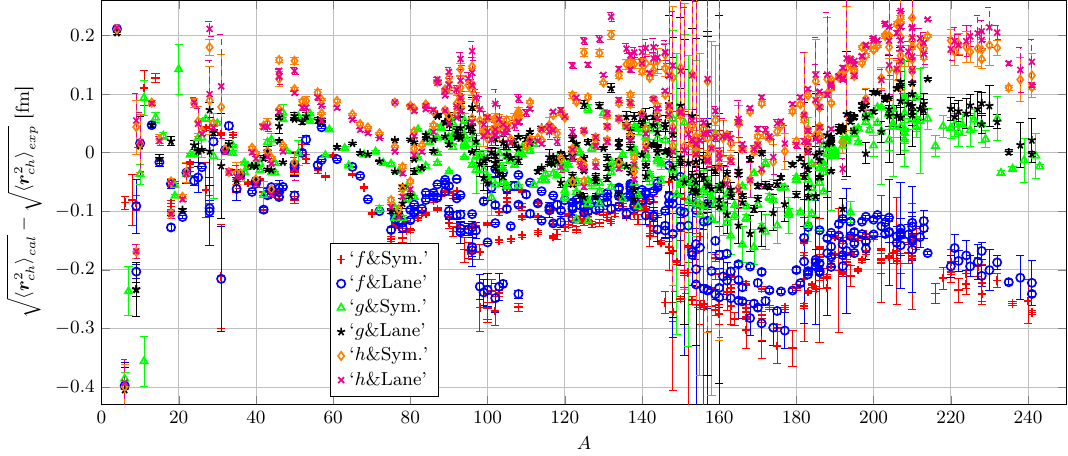}
    \caption{\label{fig2}(Color online) Deviations of calculated charge radii from experimental data\,\cite{AngMari2013}. The error bars represent the experimental errors. The labels `$f$\&Sym', `$f$\&Lane', `$g$\&Sym', `$g$\&Lane', `$h$\&Sym', and `$h$\&Lane' correspond to the models described in Table\,\ref{tab3}. } 
  \end{center} 
\end{figure*}
$$$$
It is seen from panel (a) of Fig.\,\ref{fig1} that one way to resolve this anomaly is to increase $V$ to approximately 55\,MeV. To achieve this, the $1d_\frac{5}{2}$ levels must be lowered to approximately $-4$\,MeV for $^{17}$F (protons) and $-8$\,MeV for $^{17}$O (neutrons). To gain further insight, we perform self-consistent Skyrme-HF calculations using 
a well-established parameter set
\,\cite{CHABANAT1997710,CHABANAT1998231,Bennaceur}.  
The results of this calculations for energies agree remarkably with expectations, specifically $-7.686$\,MeV for $\nu1d_\frac{5}{2}$ in $^{17}$O and $-4.0814$~MeV for $\pi1d_\frac{5}{2}$ in $^{17}$F. This finding suggests that the large discrepancy in $V$ for these nuclei between the expected values\footnote{Here, the expected $V$ values refer to those expected from the global systematics illustrated in Panel a) of Fig.\,\ref{fig1}, whereas the experimental $V$ values are those obtained by matching the Woods-Saxon eigenvalues with the experimental data.} and the experimental data is likely due to an additional effect beyond the mean-field approach. 
This suggestion is further supported by microscopic many-body calculations using the Gamow shell model\,\cite{Nicolas20} and the multiphonon approach\,\cite{PhysRevC.95.034327}. Therefore, these two data points should be excluded from constraints on WS parameters.
Throughout this paper, we use the root mean square (RMS) error as a measure of fitting quality. The RMS error for a given quantity $y$ is defined as
\begin{equation}\label{rms}
   \displaystyle \text{RMS}(y) = \sqrt{ \frac{1}{n-1} \sum_{i=1}^n \left(y_i - y^{fit}_i\right)^2 }, 
\end{equation}
where $n$ stands for sample size, $y_i$ represents the observed data points (e.g. the $V$ values obtained by matching the WS eigenvalues with experimental data), and $y^{fit}_i$ denotes the fitted values. 

For the fittings of $V$ ($y=V$), we include 38 states ($n=38$) as listed in Table\,\ref{tab1}, excluding the anomaly in the nuclei with $A=17$ discussed above. The optimal $\text{RMS}(V)$ values obtained with the `$f$\&Sym.' and `$f$\&Lane' models are 1.336\,MeV and 1.175\,MeV, respectively. This difference (reduction) in $\text{RMS}(V)$ by 161\,keV is undoubtedly attributed to the additional mass dependence of $3/A$ in the derived Lane's formula\,\eqref{Lane1}. Notably, the $V$ values for self-conjugated nuclei, where the isovector component vanishes identically, are consistent with this mass dependence, particularly in light nuclei, as evident from  panel (c) of Fig.\,\ref{fig1}. The single-particle energies for $^{12}$C, $^{20}$Ne, $^{56}$Ni (for protons) and $^{80}$Zr used in this illustration, are obtained from Skyrme-HF calculations. It should be noted that the experimental uncertainties are unavailable for most cases and thus not incorporated into the present fittings. The resulting values of $V_0$ and $\kappa$ are 
53.364 and 0.674 for the `$f$\&sym.' model, and 51.294 and 0.65 for the `$f$\&Lane' model, respectively, as reported in the first two rows of Table\,\ref{tab2}. 


To gather more information, we apply the obtained parameter sets to study charge-radius systematics. 
This study includes nearly all nuclei appeared in the charge-radius compilation in Ref.\,\cite{AngMari2013}, excluding those with $A<6$. The expectation value of the square-radius operator is treated within the closure approximation\,\cite{XaNa2018}, while the occupation numbers are obtained by assuming an equal filling of proton orbitals. 
Corrections due to the finite size of nucleons and center-of-mass motion are evaluated following the methods described in Ref.\,\cite{XaNa2018}. 
The results for the deviations of the calculated charge radii from experimental data are plotted against $A$ 
as illustrated in Fig.\,\ref{fig2}. 
It is seen that, for heavy neutron-rich nuclei, both the `$f$\&Sym.' and `$f$\&Lane' models
consistently underestimate the experimental data, with the charge-radius values
obtained with `$f$\&Lane' being slightly more accurate. 
In medium nuclei with $38\lesssim A\lesssim 62$, both models yield results that 
closely coincide and show better agreement with the experimental data. 
In lighter nuclei, however, the values calculated using `$f$\&Lane' become smaller than those obtained
using `$f$\&Sym.'. 
The RMS errors for charge radii are 0.142\,fm and 0.167\,fm for `$f$\&Lane' and `$f$\&Sym.', respectively. These numbers are consistent with the RMS$(V)$ values listed in the first two columns of Table\,\ref{tab2}. 
Our studies in this section strongly support the additional mass dependence of $3/A$, as derived from Lane's formula\,\eqref{Lane}. 

It should be noted that our calculations assume spherical symmetry and do not incorporate any corrections for correlations in open-shell nuclei\,\cite{PhysRevC.106.L011304,PhysRevLett.128.152501}. Therefore, only the global systematics of charge radii can be expected, and any local fluctuations should be excluded from the present discussions. 
Additionally, in the shell-model approach to $\delta_C$\,\cite{ToHa2008,XaNa2018}, the overall depth $V$ is readjusted case-by-case to reproduce the experimental data on proton and neutron separation energies. As a result, the difference between Eq.\,\eqref{sys} and \eqref{Lane1} discussed in this section does not influence
the calculated $\delta_C$ values, 
provided that an identical form factor is employed for both isoscalar and isovector parts.

\section{Necessity of a surface-peaked form factor}\label{sec5} 

It was argued that the real part of the optical potential should comprise a surface-peaked term since the imaginary part is found to exhibit surface-peaked behavior\,\cite{PINKSTON1965641}. This correspondence was demonstrated using the dispersive relation formalism\,\cite{Ahmad_1976,AHMAD196991}. Additionally, the isovector component of the local equivalent potential derived from the Skyrme-HF functional also shows surface-peaked behavior\,\cite{DG,XaNa2022}. Furthermore, the experimental observation of the matter density difference between $^{92}$Zr and $^{90}$Zr indicates a significant rearrangement of the magic proton core due to the addition of two neutrons\,\cite{HUANG2023138293}. All mean-field and beyond mean-field approaches failed to explain this observation, except the WS, which includes an additional surface-peaked isovector term. 
Moreover, one could argue that the conventional part of WS potential was already optimized through a global fitting using experimental data on the relevant observables in the vicinity of closed-shell nuclei\,\cite{SWV,BohrMott,WS1,WS2,WS3}; therefore, it should not be refitted. Instead, applications to open-shell nuclei should, in principle, be optimized by imposing appropriate extensions.  
\begin{table}[ht!]
\begin{center}
\caption{Models for the nuclear isoscalar and isovector components of the Woods-Saxon potential. The superscript `$s$' is added to indicate the incorporation of a surface-peaked form factor. The dependence of $V_1$ on $\pm(N-Z)/A$ is not prespecified.} 
\label{tab3} 
\begin{threeparttable}
\begin{ruledtabular}
\renewcommand{\arraystretch}{1} 
\begin{tabular}{p{6.5cm}|p{3cm}}
 Models & Labels \\
 \hline \\[-12pt]
  $\displaystyle 
  V(r)=V_0\Big[1 \pm \kappa\frac{(N-Z)}{A}\Big]f(r)$ & `$f$\&Sym.'\tnote{$\dagger$} \\[0.12in]
  $\displaystyle V(r)=V_0\Big[1 + \kappa\frac{3}{A} \pm \kappa\frac{(N-Z)}{A}\Big]f(r)$ & `$f$\&Lane'\tnote{\#} \\[0.12in]
\hline
  $\displaystyle V^s(r)=V_0f(r) + V_1 g(r)$ & `$g$\&Sym.'\tnote{\$} \\[0.12in]
  $\displaystyle V^s(r)=V_0\Big(1+\kappa\frac{3}{A}\Big)f(r) + V_1 g(r)$ & `$g$\&Lane'\tnote{\$} \\[0.12in]
  $\displaystyle V^s(r)=V_0f(r) + V_1 h(r)$ & `$h$\&Sym.'\tnote{\$} \\[0.12in]
  $\displaystyle V^s(r)=V_0\Big(1+\kappa\frac{3}{A}\Big)f(r) + V_1 h(r)$ & `$h$\&Lane'\tnote{\$} 
\end{tabular}
\end{ruledtabular}
\begin{tablenotes}
{ \raggedright
\item[$\dagger$] Original symmetry term\,\eqref{sys}, \\
\item[\#] Original Lane's formula for bound states\,\eqref{Lane1}, \\
\item[\$] Extended versions incorporating surface-peaked form factors. \\
}
\end{tablenotes}
\end{threeparttable}
\end{center}
\end{table}

In general, the radial form factors for the first and second terms in Eq.\,\eqref{sys} or \eqref{Lane1} are not necessarily identical. 
In this section, we employ the Fermi like function~\eqref{f} as a form factor for the isoscalar component, including the additional term, proportional to $3/A$, in Eq.~\eqref{Lane1}. 
For the isovector component, we consider two different surface-peaked form factors, as employed in Ref.\,\cite{ToHa2008}. The first is given by 
\begin{equation}\label{g}
\displaystyle
g(r) = -\left(\frac{\hbar}{m_\pi c}\right)^2\frac{1}{r} \frac{d}{dr}f_s(r),  
\end{equation}
where $f_s(r)$ takes the same form as $f(r)$ but with a smaller length parameter. $g(r)$ is equivalent to the spin-orbit form factor used in Refs.\,\cite{ToHa2008,XaNa2018}. 
The constant $\hbar^2/(m_\pi c)^2$ is introduced to maintain the correct dimensionality. The negative sign in Eq.\,\eqref{g} is introduced to cancel out the one resulting from the first derivative of $f(r)$. The second form factor is given by 
\begin{equation}\label{h}
\displaystyle
    h(r) = a_0^2\left[ \frac{d}{dr}f(r) \right]^2. 
\end{equation}
It is interesting to note that $h(r)$ exhibits a bell shape, very similar to a Gaussian function, as displayed in Fig.\,\eqref{fig0}. Specifically, it has a peak centered exactly at $r=R$ and monotonically decreases on both sides. The shape of $h(r)$ is symmetric with respect to its peak position. In contrast, the peak position of $g(r)$ is located at a slightly smaller distance, even if we use identical length parameter, as can been seen from Fig.\,\ref{fig0}. Additionally, $g(r)$ has a singularity at the coordinate origin due to the presence of $1/r$. The shape of $g(r)$ is similar to the spin-orbit form factor\,\cite{PhysRevC.108.064310}, and its magnitude is approximately five times larger than that of $h(r)$. Furthermore, it has been emphasized
in Ref.\,\cite{XaNa2018} that $h(r)$ leads to an undesirable quadratic correlation between the calculated charge radii and the length parameter for several cases in light-mass regions. This behavior significantly deteriorates the efficiency of the optimization processes applied to the calculations of $\delta_C$ discussed therein. Our generalized parametrizations, incorporating these surface-peaked form factors, are given in the last four rows of Table\,\ref{tab3}. 

\begin{table*}[ht!]
\begin{center}
\caption{Parameters values obtained in this work. The parameter $\kappa$ is dimensionless. The units of the other quantities, including the RMS errors of the fittings, are MeV. The complete expression of each model is given in Table\,\ref{tab3}.} 
\label{tab2} 
\begin{ruledtabular}
\begin{tabular}{c|cc|cccc|c}
 Model's label & $V_0$ & $\kappa$ & $a$ & $b$ & $c$ & $d$ & fitting RMS errors \\ 
 \hline
 `$f$\&sym.' & 53.364 & 0.674 & - &  - & - & - & RMS($V$)=1.336 \\  
 `$f$\&Lane' & 51.294 & 0.65 & - & - & - & - & RMS($V$)=1.175 \\
 \hline
 `$g$\&sym.' &  53.364 & 0.674  & $-8769.408$ & 919.462 & $-332.236$ & $-2.186 $ & RMS($V_1$)=33 \\  
 `$g$\&Lane' &  51.294 & 0.65  & $-8440.303$ & 326.361 & $-322.775$ & $ 2.919$ & RMS($V_1$)=30 \\
 `$h$\&sym.' &  53.364 & 0.674  & $-4859.172$ & $924.204$ & $-791.956$ & $-4.691$ & RMS($V_1$)=35 \\  
 `$h$\&Lane' &  51.294 & 0.65  & $-5377.055$ & $-14.536$ & $-735.791$ & $11.634$ & RMS($V_1$)=42 \\
\end{tabular}
\end{ruledtabular}
\end{center}
\end{table*}

Note that with differentiated form factors, the isoscalar and isovector strengths are no longer redundant, even for a given nucleus. Therefore, they should, in principle, be independently determined (the $V$ values obtained in Sec.\,\ref{sec4} are not applicable).
For simplicity, we fix $V_0$ and $\kappa$ (isoscalar component) to the respective values obtained in Sec.\,\ref{sec4}. As usual, the other WS parameters are kept fixed at their standard values\,\cite{XaNa2018}. Therefore, we adjust only $V_1$ (see Table\,\ref{tab3} for the complete expressions) to reproduce the experimental energies provided 
in the fifth column of Table\,\ref{tab1}. The resulting $V_1$ values are visualized against $\pm(N-Z)/A$ in  panel (d) of Fig.\,\ref{fig1}. The corresponding numerical values are given in the last four columns of Table\,\ref{tab1}. 
Note that the $V_1$ values depend on specific models for the isoscalar and isovector components of the potential. 
It is seen from Fig.\,\ref{fig1} that $V_1$ does not vary linearly with $\pm(N-Z)/A$; 
instead, it shows a much stronger correlation when higher-order terms in $\pm(N-Z)/A$ are included. 
We emphasize that these higher-order effects are not apparent with the `$f$\&Sym.' and `$f$\&Lane' models employed in Sect.\,\ref{sec4}. We find that it is sufficient to include up to the third order term, namely 
\begin{equation}\label{higher}
   V_1 = a\cdot I^3 + b\cdot I^2 + c\cdot I + d, 
\end{equation}
where $I=\pm(N-Z)/A$. The coefficients $a$, $b$, $c$ and $d$ are expected to be constant or nucleus-independent. 

Once again, we employ the RMS error, denoted as RMS($V_1$), to measure the fitting quality of the 
$V_1$ values to Eq.\,\eqref{higher}. 
We find that $h(r)$ produces a slightly larger RMS($V_1$) compared to $g(r)$ in all cases. Among the models listed in the last four rows of Table\,\ref{tab3}, the one labeled `$g$\&Lane' yields the smallest RMS($V_1$) value, specifically 30\,MeV. 
Therefore, these results support the additional mass dependence derived from Lane's formula\,\eqref{Lane}, as emphasized in Sec.\,\ref{sec4}, and also endorse $g(r)$ as an appropriate form factor for the isovector component.  

The numerical values of RMS($V_1$) and the coefficients in Eq\,\eqref{higher} for the different models are provided in the last four rows of Table\,\ref{tab2}. Note that these RMS($V_1$) values should not be compared with the RMS($V$) values obtained for the fittings in Sec.\,\ref{sec4}. 
Additionally, the $V_1$ values for the $\pi3s_\frac{1}{2}$ states in $^{207}$Tl and $^{208}$Pb are excluded due to larger deviations from the global systematics when incorporating a surface-peaked form factor, as can be seen from panel (d) of Fig.\,\ref{fig1}. This issue arises because these states lack a centrifugal barrier and are subsequently highly sensitive to small deficiencies in the models. 


It is noticeable that although our fittings in this section impose the conventional form factor for the isoscalar terms, if a surface peaking is also favorable for this isoscalar component, its contribution would be captured by the last term on the right-hand side in Eq.\,\eqref{higher}. However, our results do not provide any evidence of such behavior, as none of the curves depicted 
in  panel (d) of Fig.\,\ref{fig1} show a considerable intercept. 

To further justify the necessity of a surface-peaked form factor, 
we perform a systematic study of charge radii similar to that in Sec.\,\ref{sec4}, employing the new parameter sets from the last four rows of Table\,\ref{tab3}.  
The resulting deviations between the calculated and experimental values are plotted against $A$ as displayed in Fig.\,\ref{fig2}. 
These results indicate that both  $g(r)$ and $h(r)$ are effectively repulsive, 
increasing charge radii, especially in neutron-rich nuclei.  
Remarkably, the charge-radius values obtained with $h(r)$ consistently overestimate the experimental data, in contrast to those produced with $f(r)$. 
However,  the incorporation of $g(r)$ leads to a significant improvement to the charge-radius systematics, with values falling between those obtained with $h(r)$ and $f(r)$. The RMS errors for the calculated charge radii obtained in the current section are 0.062\,fm (`$g$\&Lane'), 0.066\,fm (`$g$\&Sym.'), 0.103\,fm (`$h$\&Lane'), and 0.109\,fm (`$h$\&Sym.'). 

To complete our analysis, 
we show the numerical results for energies of the single-particle/hole states selected for the fittings of $V$ and $V_1$. 
Similar to charge radii, these calculations employ the parameter sets obtained in Table\,\ref{tab3}. The results further demonstrate that the derived Lane's formula\,\eqref{Lane1} is more accurate than the symmetry term\,\eqref{sys}, as evidenced by the RMS errors [denoted as RMS($E$)] provided in the last row of Table\,\ref{tab2x}. However, the incorporation of a surface-peaked form factor tends to increase the RMS($E$) in these nuclei. Based on all the results obtained in this study, we conclude that a surface-peaked form factor is necessary only for nuclei in the regions far from the valley of stability, including those relevant to the fundamental physics studies in Ref.\,\cite{HaTo2020}. 

\section{Summary}\label{sec6}

In this work, we presents a detailed study of the isovector component within the phenomenological Woods-Saxon potential. Lane's isospin dependence, originally designed for the nuclear optical model, is scrutinized in the context of bound nucleon states. We demonstrate that this formulation for bound nuclei in the ground states slightly deviates from the conventional symmetry term.
When expressed in terms of parameters associated with the compound nucleus, Lane's formula appears as a sum of the conventional symmetry term and a residual isoscalar term proportional to $3/A$. 
We also perform a comparative analysis between these different approaches. 
Our results indicate that the inclusion of $3/A$ gives rise to a slightly improved correlation in fitting for the overall depth $V$. 
Lane's formula also provides a more accurate description of charge-radius systematics across the nuclear chart compared to the conventional parametrization. 
Furthermore, we explore two distinct surface-peaked form factors: one involving the first derivative of the Fermi function divided by
the radial coordinate, similar to the usual spin-orbit form factor, and another involving the squares of the first derivative of the Fermi function. Among these variants, the spin-orbit-like form factor, when combined with Lane's formula, shows a remarkably better accuracy. To further validate the necessity of surface peaking, we apply the obtained parameter sets to calculate charge radii throughout the nuclear landscape. The results of these calculations consistently favor the spin-orbit-like form factor and Lane's isospin dependence. The findings from the present research have immediate implications for the calculations of the isospin-symmetry breaking correction using the phenomenological shell model, which currently yields the greatest consistency with the standard model. 
Our results suggest that the correction values previously obtained without an appropriate surface-peaked form factor should be rejected. This rejection will reduce the uncertainty in the theoretical inputs for the subsequent tests of the standard model. 

\begin{acknowledgments} 
L. Xayavong and Y. Lim are supported by the National Research Foundation of Korea(NRF) grant funded by the Korea government(MSIT)(No. 2021R1A2C2094378). Y. Lim is also supported by the Yonsei University Research Fund of 2024-22-0121. 
N. A. Smirnova acknowledges the financial support of CNRS/IN2P3, France, via ENFIA Master project.
\end{acknowledgments}
%

%

\appendix*
\section{Supplemental numerical data} 

\begin{table*}[ht!]
\caption{Data used for the fittings conducted in this work. Energies are listed in the fifth column,
with values for closed-(sub)shell nuclei taken from Ref.\,\cite{Vautherin}, those for closed-(sub)shell nuclei with an additional nucleon/hole from Ref.\,\cite{SWV}, 
and those for neutrons in $^{50}$Ti, $^{52}$Cr, $^{54}$Fe, and $^{56}$Ni 
from Ref.\,\cite{Bespalova2007}. The $V$ values are provided in the sixth column. 
The $V_1$ values for the various models in Table\,\ref{tab3} are listed in the subsequent columns. The units of $E$, $V$ and $V_1$ are MeV.} 
\label{tab1} 
\begin{threeparttable}
\begin{ruledtabular}
\begin{tabular}{c|c|c|c|c|c|c|c|c|c} 
Nuclei	&	$\pm(N-Z)/A$	&	$\pm(N-Z)/A+3/A$	&	States	&	$E$	&	$V$	&	$V_1$\tnote{$\dagger$}	&	$V_1$\tnote{\#}	&	$V_1$\tnote{\$}	&	$V_1$\tnote{\&}	\\
\hline		
$^{15}$N	&	0.067	&	0.267		&	$\pi 1p\frac{1}{2}$ (hole)	&	-12.130	&	57.710	&	-21.980	&	-67.280	&	1.300	&	3.800	\\
$^{15}$O	&	0.067	&	0.267	&	$\nu 1p_\frac{1}{2}$ (hole)	&	-15.660	&	58.390	&	-25.630	&	-79.080	&	-7.000	&	-2.100	\\
$^{16}$O	&	0.000	&	0.188		&	$\pi 1p_\frac{1}{2}$ (particle)	&	-12.100	&	56.510	&	-16.790	&	-51.000	&	-26.600	&	-82.300	\\
$^{16}$O	&	0.000	&	0.188		&	$\nu1p_\frac{1}{2}$ (particle)	&	-15.700	&	56.570	&	-17.240	&	-52.900	&	37.800	&	119.700	\\
$^{17}$F	&	-0.059	&	0.118		&	$\pi1d_\frac{5}{2}$ (particle)	&	-0.600	&	48.360	&	29.260	&	75.670	&	51.700	&	134.700	\\
$^{17}$O	&	-0.059	&	0.118		&	$\nu1d_\frac{5}{2}$ (particle)	&	-4.140	&	48.330	&	29.420	&	76.580	&	52.100	&	136.600	\\
$^{39}$K	&	0.026	&	0.103		&	$\pi1d_\frac{3}{2}$ (hole)	&	-8.330	&	54.600	&	-11.230	&	-25.360	&	-6.700	&	-15.300	\\
$^{39}$Ca	&	0.026	&	0.103		&	$\nu1d_\frac{3}{2}$ (hole)	&	-15.640	&	54.920	&	-14.490	&	-33.350	&	-9.800	&	-22.900	\\
$^{40}$Ca	&	0.000	&	0.075		&	$\pi1d_\frac{3}{2}$ (particle)	&	-8.300	&	54.340	&	-9.060	&	-20.410	&	-5.000	&	-11.500	\\
$^{40}$Ca	&	0.000	&	0.075			&	$\nu1d_\frac{3}{2}$ (particle)	&	-15.600	&	54.150	&	-7.590	&	-17.490	&	-3.400	&	-8.000	\\
$^{41}$Sc	&	-0.024	&	0.049		&	$\pi1f_\frac{7}{2}$ (particle)	&	-1.090	&	50.481	&	27.400	&	54.520	&	31.000	&	61.700	\\
$^{41}$Ca	&	-0.024	&	0.049		&	$\nu 1f_\frac{7}{2}$ (particle)	&	-8.360	&	50.500	&	27.240	&	54.960	&	30.900	&	62.400	\\
$^{47}$K	&	0.191	&	0.255		&	$\pi1d_\frac{3}{2}$ (hole)	&	-16.180	&	60.490	&	-76.940	&	-163.690	&	-76.200	&	-162.400	\\
$^{47}$Ca	&	-0.149	&	-0.085			&	$\nu1f_\frac{7}{2}$ (hole)	&	-10.000	&	49.410	&	41.380	&	82.060	&	42.100	&	83.400	\\
$^{48}$Ca	&	0.167	&	0.229			&	$\pi1d_\frac{3}{2}$ (particle)	&	-15.700	&	59.750	&	-70.250	&	-149.320	&	-70.000	&	-148.900	\\
$^{48}$Ca	&	-0.167	&	-0.104		&	$\nu1f_\frac{7}{2}$ (particle)	&	-9.940	&	48.820	&	48.220	&	95.530	&	48.500	&	96.000	\\
$^{49}$Sc	&	0.143	&	0.204			&	$\pi1f_\frac{7}{2}$ (particle)	&	-9.350	&	57.060	&	-39.800	&	-75.750	&	-40.000	&	-76.200	\\
$^{49}$Ca	&	-0.184	&	-0.122		&	$\nu2p_\frac{3}{2}$ (particle)	&	-4.600	&	45.440	&	103.100	&	196.040	&	102.900	&	195.500	\\
$^{50}$Ti	&	-0.120	&	-0.060			&	$\nu1f_\frac{7}{2}$ (particle)	&	-11.590	&	50.100	&	35.800	&	70.200	&	35.100	&	68.700	\\
$^{52}$Cr	&	-0.077	&	-0.019		&	$\nu1f_\frac{7}{2}$ (particle)	&	-12.510	&	50.420	&	33.400	&	64.900	&	31.700	&	61.600	\\
$^{54}$F	&	-0.037	&	0.019		&	$\nu1f_\frac{7}{2}$ (particle)	&	-14.970	&	52.770	&	6.900	&	13.200	&	4.400	&	8.300	\\
$^{56}$Ni	&	0.000	&	0.054		&	$\nu1f_\frac{7}{2}$ (particle)	&	-16.650	&	54.090	&	-8.700	&	-16.700	&	-12.100	&	-23.200	\\
$^{55}$Co	&	0.018	&	0.073	&	$\pi1f_\frac{7}{2}$ (hole)	&	-7.170	&	53.830	&	-5.480	&	-10.290	&	-8.200	&	-15.500	\\
$^{55}$Ni	&	0.018	&	0.073		&	$\nu1f_\frac{7}{2}$ (hole)	&	-16.640	&	54.500	&	-13.530	&	-25.770	&	-16.400	&	-31.200	\\
$^{90}$Zr	&	0.111	&	0.144	&	$\pi2p_\frac{1}{2}$ (particle)	&	-7.030	&	55.850	&	-50.320	&	-76.410	&	-70.000	&	-106.600	\\
$^{90}$Zr	&	-0.111	&	-0.078	&	$\nu1g_\frac{9}{2}$ (particle)	&	-12.000	&	49.360	&	63.760	&	105.320	&	49.100	&	80.300	\\
$^{57}$Cu	&	-0.018	&	0.035	&	$\pi2p_\frac{3}{2}$ (particle)	&	-0.690	&	50.950	&	37.990	&	69.270	&	33.500	&	61.100	\\
$^{57}$Ni	&	-0.018	&	0.035	&	$\nu2p_\frac{3}{4}$ (particle)	&	-10.250	&	51.170	&	32.660	&	58.710	&	28.100	&	50.500	\\
$^{100}$Sn	&	0.000	&	0.030	&	$\pi1g_\frac{9}{2}$ (particle)	&	-2.920	&	52.650	&	11.200	&	17.300	&	-5.500	&	-8.600	\\
$^{100}$Sn	&	0.000	&	0.030	&	$\nu1g_\frac{9}{2}$ (particle)	&	-17.930	&	54.030	&	-10.900	&	-17.300	&	-28.300	&	-44.400	\\
$^{131}$In	&	0.252	&	0.275	& $\pi1g_\frac{9}{2}$ (hole)	&	-15.710	&	60.580	&	-138.000	&	-193.200	&	-161.600	&	-223.800	\\
$^{131}$Sn	&	-0.237	&	-0.214	&	$\nu2d_\frac{3}{2}$ (hole)	&	-7.310	&	43.380	&	221.330	&	306.200	&	192.400	&	262.300	\\
$^{133}$Sb	&	0.233	&	0.256	&	$\pi1g_\frac{7}{2}$ (particle)	&	-9.680	&	61.610	&	-150.900	&	-216.100	&	-172.800	&	-244.200	\\
$^{133}$Sn	&	-0.248	&	-0.226	&	$\nu2f_\frac{7}{2}$	(particle) &	-2.470	&	44.050	&	207.200	&	274.800	&	187.300	&	249.600	\\
$^{207}$Tl	&	0.217	&	0.232	&	$\pi3s_\frac{1}{2}$ (hole)	&	-8.010	&	60.300	&	-316.000	&	-399.100	&	-399.100	&	-534.600	\\
$^{207}$Pb	&	-0.208	&	-0.193	& $\nu3p_\frac{1}{2}$ (hole)	&	-7.370	&	44.760	&	276.700	&	310.700	&	228.400	&	255.500	\\
$^{209}$Bi	&	0.206	&	0.220	&	$\pi1h_\frac{9}{2}$ (particle)	&	-3.800	&	60.460	&	-160.800	&	-196.200	&	-193.200	&	-231.900	\\
$^{209}$Pb	&	-0.215	&	-0.201	&	$\nu2g_\frac{9}{2}$ (particle)	&	-3.940	&	45.250	&	248.400	&	279.600	&	201.800	&	227.800	\\
$^{208}$Pb	&	0.212	&	0.226	&	$\pi3s_\frac{1}{2}$ (particle)	&	-8.030	&	60.480	&	-326.700	&	-414.400	&	-410.800	&	-552.700	\\
$^{208}$Pb	&	-0.212	&	-0.197	&	$\nu3p_\frac{1}{2}$ (particle)	&	-7.380	&	44.670	&	280.500	&	314.700	&	232.000	&	259.200	\\
\end{tabular}
\end{ruledtabular}
\begin{tablenotes}
{ \raggedright
\item[$\dagger$] Evaluated with $g(r)$, while the remaining part is fixed with the parameter set `$f$\&Sym.', \\
\item[\#] Evaluated with $h(r)$, while the remaining part is fixed with the parameter set `$f$\&Sym.', \\
\item[\$] Evaluated with $g(r)$, while the remaining part is fixed with the parameter set `$f$\&Lane', \\
\item[\&] Evaluated with $h(r)$, while the remaining part is fixed with the parameter set `$f$\&Lane'. \\
}
\end{tablenotes}
\end{threeparttable}
\end{table*}

\begin{table*}[ht!]
\begin{center}
\caption{Single-particle/hole energies obtained with the Woods-Saxon parameter sets produced in this work. All values are in MeV. The RMS errors are provided in the last row. The experimental values are given in the third column.} 
\label{tab2x} 
\begin{ruledtabular}
\begin{tabular}{c|c|c|ccccccc}
Nuclei	& States &  $E$ & 	`$f$\&Sym'	&	`$f$\&Lane'	&	`$g$\&Sym.'	&	`$g$\&Lane'	&	`$h$\&Sym.'	&	`$h$\&Lane'	\\
\hline
$^{15}$N	&	$\pi1p\frac{1}{2}$	(hole) 	&	-12.130	&	-11.012	&	-13.571	&	-12.237	&	-14.529	&	-11.516	&	-13.572	\\
$^{15}$O	&	$\nu1p_\frac{1}{2}$	(hole) 	&	-15.660	&	-14.123	&	-16.726	&	-15.349	&	-17.682	&	-14.589	&	-16.696	\\
$^{16}$O	&	$\pi1p_\frac{1}{2}$	(particle) 	&	-12.100	&	-10.261	&	-12.711	&	-10.500	&	-12.386	&	-10.414	&	-12.340	\\
$^{16}$O	&	$\nu1p_\frac{1}{2}$	(particle) 	&	-15.700	&	-13.792	&	-16.290	&	-14.034	&	-15.963	&	-13.943	&	-15.923	\\
$^{39}$K	&	$\pi1d_\frac{3}{2}$	(hole) 	&	-8.330	&	-8.122	&	-8.403	&	-8.267	&	-8.228	&	-8.181	&	-8.038	\\
$^{39}$Ca	&	$\nu1d_\frac{3}{2}$	(hole) 	&	-15.640	&	-15.210	&	-15.499	&	-15.337	&	-15.307	&	-15.222	&	-15.108	\\
$^{40}$Ca	&	$\pi1d_\frac{3}{2}$	(particle) 	&	-8.300	&	-7.658	&	-7.941	&	-7.815	&	-7.731	&	-7.781	&	-7.638	\\
$^{40}$Ca	&	$\nu1d_\frac{3}{2}$	(particle) 	&	-15.600	&	-15.061	&	-15.355	&	-15.219	&	-15.144	&	-15.180	&	-15.064	\\
$^{41}$Sc	&	$\pi1f_\frac{7}{2}$	(particle) 	&	-1.090	&	-2.472	&	-2.766	&	-2.591	&	-2.516	&	-2.564	&	-2.345	\\
$^{41}$Ca	&	$\nu1f_\frac{7}{2}$	(particle) 	&	-8.360	&	-9.753	&	-10.060	&	-9.887	&	-9.816	&	-9.879	&	-9.678	\\
$^{47}$K	&	$\pi1d_\frac{3}{2}$	(hole) 	&	-16.180	&	-16.008	&	-15.683	&	-17.323	&	-18.215	&	-15.037	&	-15.379	\\
$^{47}$Ca	&	$\nu1f_\frac{7}{2}$	(hole) 	&	-10.000	&	-8.997	&	-9.318	&	-6.215	&	-6.973	&	-8.542	&	-8.896	\\
$^{48}$Ca	&	$\pi1d_\frac{3}{2}$	(particle) 	&	-15.700	&	-15.414	&	-15.104	&	-15.870	&	-16.432	&	-14.399	&	-14.483	\\
$^{48}$Ca	&	$\nu1f_\frac{7}{2}$	(particle) 	&	-9.940	&	-8.899	&	-9.222	&	-5.140	&	-6.114	&	-8.293	&	-8.752	\\
$^{49}$Sc	&	$\pi1f_\frac{7}{2}$	(particle) 	&	-9.350	&	-10.440	&	-10.134	&	-10.521	&	-10.828	&	-10.221	&	-10.079	\\
$^{49}$Ca	&	$\nu2p_\frac{3}{2}$	(particle) 	&	-4.600	&	-5.320	&	-5.574	&	-2.876	&	-3.635	&	-4.094	&	-4.636	\\
$^{50}$Ti	&	$\nu1f_\frac{7}{2}$	(particle) 	&	-11.590	&	-10.813	&	-10.994	&	-9.548	&	-9.841	&	-10.850	&	-10.879	\\
$^{52}$Cr	&	$\nu1f_\frac{7}{2}$	(particle) 	&	-12.510	&	-12.647	&	-12.689	&	-12.547	&	-12.387	&	-12.942	&	-12.644	\\
$^{54}$Fe	&	$\nu1f_\frac{7}{2}$	(particle) 	&	-14.970	&	-14.402	&	-14.310	&	-14.647	&	-14.220	&	-14.696	&	-14.167	\\
$^{56}$Ni	&	$\nu1f_\frac{7}{2}$	(particle) 	&	-16.650	&	-16.081	&	-15.859	&	-16.223	&	-15.669	&	-16.211	&	-15.537	\\
$^{55}$Co	&	$\pi1f_\frac{7}{2}$	(hole) 	&	-7.170	&	-7.314	&	-7.088	&	-7.342	&	-6.820	&	-7.374	&	-6.683	\\
$^{55}$Ni	&	$\nu1f_\frac{7}{2}$	(hole) 	&	-16.640	&	-16.268	&	-16.035	&	-16.283	&	-15.753	&	-16.290	&	-15.613	\\
$^{90}$Zr	&	$\pi2p_\frac{1}{2}$	(particle) 	&	-7.030	&	-8.145	&	-7.224	&	-6.665	&	-5.999	&	-7.234	&	-6.322	\\
$^{90}$Zr	&	$\nu1g_\frac{9}{2}$	(particle) 	&	-12.000	&	-12.176	&	-11.647	&	-12.306	&	-11.725	&	-12.938	&	-12.156	\\
$^{57}$Cu	&	$\pi2p_\frac{3}{2}$	(particle) 	&	-0.690	&	-1.868	&	-1.701	&	-2.098	&	-1.706	&	-2.028	&	-1.462	\\
$^{57}$Ni	&	$\nu2p_\frac{3}{4}$	(particle) 	&	-10.250	&	-11.290	&	-11.109	&	-11.538	&	-11.115	&	-11.473	&	-10.876	\\
$^{100}$Sn	&	$\pi1g_\frac{9}{2}$	(particle) 	&	-2.920	&	-3.493	&	-2.640	&	-3.604	&	-2.492	&	-3.614	&	-2.340	\\
$^{100}$Sn	&	$\nu1g_\frac{9}{2}$	(particle) 	&	-17.930	&	-17.382	&	-16.496	&	-17.492	&	-16.349	&	-17.493	&	-16.220	\\
$^{131}$In	&	$\pi1g_\frac{9}{2}$	(hole) 	&	-15.710	&	-17.319	&	-15.605	&	-17.085	&	-17.157	&	-14.940	&	-14.987	\\
$^{131}$Sn	&	$\nu2d_\frac{3}{2}$	(hole) 	&	-7.310	&	-8.310	&	-7.841	&	-6.578	&	-6.762	&	-8.982	&	-8.607	\\
$^{133}$Sb	&	$\pi1g_\frac{7}{2}$	(particle) 	&	-9.680	&	-9.787	&	-8.273	&	-9.241	&	-9.186	&	-7.630	&	-7.467	\\
$^{133}$Sn	&	$\nu2f_\frac{7}{2}$	(particle)	&	-2.470	&	-2.708	&	-2.297	&	-0.929	&	-1.223	&	-1.721	&	-1.737	\\
$^{207}$Pb	&	$\nu3p_\frac{1}{2}$	(hole) 	&	-7.370	&	-8.154	&	-7.430	&	-9.525	&	-8.951	&	-9.306	&	-8.557	\\
$^{209}$Bi	&	$\pi1h_\frac{9}{2}$	(particle) 	&	-3.800	&	-4.046	&	-2.342	&	-1.919	&	-1.264	&	-1.512	&	-0.781	\\
$^{209}$Pb	&	$\nu2g_\frac{9}{2}$	(particle) 	&	-3.940	&	-4.204	&	-3.481	&	-5.170	&	-4.680	&	-4.932	&	-4.274	\\
$^{208}$Pb	&	$\nu3p_\frac{1}{2}$	(particle) 	&	-7.380	&	-8.138	&	-7.418	&	-9.458	&	-8.908	&	-9.303	&	-8.565	\\
\hline
\multicolumn{3}{c|}{RMS($E$)}		&	0.941	&	0.832	&	1.55	&	1.47	&	1.204	&	1.099	\\
\end{tabular}
\end{ruledtabular}
\end{center}
\end{table*}

\end{document}